\documentstyle[aps]{revtex}

\begin{document}

\title{Diffusion in Stationary Flow from Mesoscopic Non-equilibrium Thermodynamics}

\author{I. Santamar\'{\i}a-Holek, D. Reguera, and J. M. Rub\'{\i} }

\address{Departament de F\'{\i}sica Fonamental-CER Física de Sistemes Complexos\\
 Facultat de F\'{\i}sica\\
 Universitat de Barcelona\\
 Diagonal 647, 08028 Barcelona, Spain\\
 }

\maketitle
\begin{abstract}
We analyze the diffusion of a Brownian particle in a fluid under stationary
flow. By using the scheme of non-equilibrium thermodynamics in phase space,
we obtain the Fokker-Planck equation which is compared with others derived from
kinetic theory and projector operator techniques. That equation exhibits violation
of the fluctuation dissipation-theorem. By implementing the hydrodynamic regime
described by the first moments of the non-equilibrium distribution, we find
relaxation equations for the diffusion current and pressure tensor,
allowing us to arrive at a complete description of the system in the inertial
and diffusion regimes. The simplicity and generality of the method we propose,
makes it applicable to more complex situations, often encountered in problems
of soft condensed matter, in which not only one but more degrees of freedom
are coupled to a non-equilibrium bath.

\end{abstract}

\section{INTRODUCTION}

Liquid matter when subjected to the action of external forces or gradients exhibits
peculiar characteristics which do not manifest in the absence of those external
inputs. Its statistical mechanical properties present significant features as
the appearance of long-range correlations and phase transitions, or the violation
of the fluctuation-dissipation theorem (see for example Refs. \cite{lenan}-\cite{onuki}).
The aim of non-equilibrium statistical mechanical theories is precisely to explain
the behavior of systems in such situations, away from equilibrium.

These characteristics, far from being specific of pure substances as simple
fluids \cite{dorfman}, also manifest in complex fluids \cite{fragile}. Typical
situations of transport in liquids or liquid-like systems involve the joint
motion of liquid and solid or solid-like phases under the action of an external
input. Many examples can be found in systems as polymers \cite{Doi}, suspensions
of neutral \cite{vandeven} and field-responsive particles \cite{vilar} and
granular media under shear flow \cite{jaeger}. The presence of shear flow significantly
modifies transport properties, and may induce the appearance of new phases which
otherwise would remain hidden (see for instance Refs. \cite{chen}-\cite{yamamoto}).
That is the reason why the influence of the shear flow in the dynamics is a
subject which has received much attention in the last years, specially in the
domain of soft condensed matter. 

Our purpose in this paper is precisely to analyze one of the simplest examples
of a system coupled to a non-equilibrium bath, whose
physical realization is a set of non-interacting Brownian particles moving in
a fluid in stationary flow. The suspended objects diffuse in and
are convected by the fluid; their motion may eventually be influenced by the
presence of external fields. These are the basic ingredients controlling the
dynamics of the suspended phase. Since the Brownian objects are of mesoscopic
nature, the dynamic description demands a mesoscopic treatment in terms of a
probability distribution function. The evolution in time of this quantity is
governed by Fokker-Planck and Smoluchowski equations. These equations constitute
the basis for a mesoscopic description of the system and enable one to extract
macroscopic information from the evolution equations for the moments of the
distribution. It becomes then of primary importance to establish simple methods
able to provide expressions of those equations in situations outside equilibrium.

Kinetic equations of the Fokker-Planck type have been basically derived from
kinetic theory of gases using the diffusion approximation in the Boltzmann equation
\cite{lenan,physicalkietics,Zubarev}, from the theory of stochastic processes
through the master equation \cite{kampen} or by means of projection operator
techniques  \cite{kampi-oppi,deutch}. It has also been shown that those
kinetic equations can be derived from mesoscopic non-equilibrium thermodynamics
(MNET)  \cite{degroot}-\cite{hydrodynamic-fluctuations}. As in non-equilibrium
thermodynamics, the basic point of this theory consists of assuming local equilibrium,
which is performed at a more basic level: at mesoscopic level. This fact enables
one to formulate a  Gibbs equation in which the  entropy, in
accordance with the concept of Gibbs entropy, also depends on a `density': the
probability density. By applying the rules of non-equilibrium thermodynamics
one obtains the entropy production and from it the corresponding linear laws
between fluxes and forces. When using these laws in the continuity (balance)
equation for the probability density one obtains the Fokker-Planck equation,
\cite{degroot,temp}. This is precisely the scheme we will adopt in this paper
to analyze diffusion in stationary flow in both, the kinetic and the hydrodynamic
regimes.

The paper is organized as follows. In Section II, we derive the Fokker-Planck
equation describing Brownian motion under the presence of an arbitrary steady
flow, in the framework of MNET. In  Section  III, we discuss
the hydrodynamic description by constructing the hierarchy of moments of the
distribution function, accounting for the macroscopic evolution of the system.
These equations reveal the presence of inertial and diffusion regimes for the
dynamics of the particles. In Section IV, we analyze the particular
case of a shear flow. The corresponding Fokker-Planck equation is compared with
the one obtained in \cite{rodriguez} by means of kinetic theory. In the diffusion
regime, we give explicit expressions for the pressure tensor and the viscosity.
Finally, in the discussion, we summarize our main results and indicate potential
applications of the formalism we have established to soft condensed matter systems.

\section{FOKKER-PLANCK DYNAMICS FROM MESOSCOPIC NON-EQUILIBRIUM THERMODYNAMICS}

We consider a dilute suspension of spherical particles of mass \( m \), immersed
in a liquid phase with constant density which acts as a heat bath. The whole
system is subjected to conditions creating a stationary flow described by the
velocity field

\begin{equation}
\label{perfil velocidad}
\vec{v}=\vec{v}_{0}(\vec{r}).
\end{equation}

Since our main purpose is to analyze the effect of the velocity gradient on
the dynamics of the particle, we will consider isothermal conditions neglecting
`viscous heating'. 

The mesoscopic nature of the suspended phase makes it necessary to analyze its
dynamics by means of a Fokker-Planck equation accounting for the evolution of
the distribution function, which may, in general, depend on the coordinates
and momenta necessary to specify the state of the suspended objects. Since we
assume no direct interactions between particles, the Brownian `gas' will be
described by means of the single-particle distribution function \( f(\vec{r},\vec{u},t) \),
which depends explicitly on the position \( \vec{r} \) , particle velocity
\( \vec{u} \), and time \( t \).

The first step towards the obtention of the Fokker-Planck equation, is the formulation
of the conservation laws for the `gas' of suspended particles. In the absence
of external body forces, the distribution function obeys the continuity equation 

\begin{equation}
\label{ec continuidad}
\frac{\partial f}{\partial t}+\nabla \cdot \vec{u}f=-\frac{\partial }{\partial \vec{u}}\cdot \vec{J}_{\vec{u}},
\end{equation}
 which introduces the current \( \vec{J}_{\vec{u}} \) in phase space. The average
of Eq.(\ref{ec continuidad}) with respect to the particle velocity \( \vec{u} \),
leads to the macroscopic equation for the balance of mass, which can be written
in the form 

\begin{equation}
\label{balance de masa1}
\frac{d\rho }{dt}=-\rho \nabla \cdot \vec{v}.
\end{equation}
 Here \( \rho (\vec{r},t) \) is the density of the Brownian `gas', given by

\begin{equation}
\label{definicion densidad}
\rho (\vec{r},t)=m\int fd\vec{u},
\end{equation}
 \( \vec{v}(\vec{r},t) \)  is the average velocity of the Brownian
particles defined through the expression 

\begin{equation}
\label{definicion flujo difusion}
\rho \vec{v}(\vec{r},t)=m\int \vec{u}fd\vec{u}.
\end{equation}
and we have defined the total derivative as  \begin{equation}
\label{derivada total}
\frac{d}{dt}\equiv \frac{\partial }{\partial t}+\vec{v}\cdot \nabla .
\end{equation}

Our purpose is to obtain the equation governing the evolution of the probability
density \( f \), therefore we need to find out the explicit expression for
the current \( \vec{J}_{\vec{u}} \). To this end, we will assume local equilibrium
for which, entropy variations are given through the Gibbs equation  \cite{degroot,temp}

\begin{equation}
\label{ec. gibbs}
\delta s=\frac{1}{T}\delta e+\frac{1}{T}p\delta \rho ^{-1}-m\int \frac{\mu }{T}\delta c_{\vec{u}}d\vec{u}.
\end{equation}
 Here \( s(\vec{r},t) \) and \( e(\vec{r},t) \) are the entropy and total
energy per unit of mass of the Brownian particles, respectively, \( p(\vec{r},t) \)
is the hydrostatic pressure, \( \mu (\vec{u},\vec{r},t) \) is the non-equilibrium
chemical potential per unit of mass, and \( c_{\vec{u}}=\frac{f}{\rho } \)
is the Brownian mass fraction \cite{degroot}. Notice that the term including
the chemical potential in Eq.(\ref{ec. gibbs}) is reminiscent of the corresponding
one for a mixture in which the different components would be specified by the
continuum `index' \( \vec{u} \). Following  the scheme of non-equilibrium
thermodynamics \cite{degroot}, we will assume that Eq.(\ref{ec. gibbs}) remains
valid for changes in time and position into a mass element followed along the
center of gravity motion of the Brownian `gas'.

Since the particles undergo only translational motion, and in view of Eqs.(\ref{ec continuidad}),
(\ref{balance de masa1}) and (\ref{ec. gibbs}), the remaining conservation
law is the one for the energy. The presence of the external flow is responsible
for the appearance in that equation of the term of `viscous heating', giving
rise to variations in the temperature field. To maintain isothermal conditions,
and following previous ideas introduced in the implementation of the so-called
`homogeneous shear'  \cite{hover,keizer}, we will assume the existence
of a local heat source capable to remove the heat generated in the process.
Under this assumption, the energy of the volume elements remains constant along
their paths and its balance equation can be omitted in the subsequent analysis.

The expression of the non-equilibrium chemical potential can be found through
the Gibbs entropy postulate for the Brownian particles written as

\begin{equation}
\label{postulado de gibbs}
s=-k_{B}\int c_{\vec{u}}ln\frac{f}{f^{l.eq}}d\vec{u}+s^{l.eq}.
\end{equation}

\noindent Here \( k_{B} \) is Boltzmann's constant and \( f^{l.eq} \) is the
local equilibrium distribution function corresponding to the reference state
described  by the local Maxwellian with respect to the stationary convective
flow \cite{lenan},

\begin{equation}
\label{funcion equilibrio local}
f^{l.eq}(\vec{u},\vec{r},t)=e^{\frac{m}{kT}[\mu _{B}-\frac{1}{2}(\vec{u}-\vec{v}_{0})^{2}]}.
\end{equation}

\noindent In this expression \( \mu _{B}(\vec{r},t) \) is the local equilibrium
chemical potential of the Brownian particles. The local equilibrium entropy
for the Brownian particles per unit of mass is given by

\begin{equation}
\label{entropia eq. local}
s^{l.eq}=-\frac{m}{T}\int c_{\vec{u}}\mu _{B}d\vec{u}+\frac{1}{T}e.
\end{equation}

After applying the total derivative in Eqs. (\ref{ec. gibbs}), (\ref{postulado de gibbs})
and (\ref{entropia eq. local}), and comparing the corresponding expressions
for the time variations of the entropy, we arrive at the desired expression
for the non-equilibrium chemical potential 

\begin{equation}
\label{potencial quimico}
\mu =\frac{k_{B}T}{m}lnf+\frac{1}{2}(\vec{u}-\vec{v}_{0})^{2}.
\end{equation}
 This expression, together with the equation for the Brownian mass fraction 

\begin{equation}
\label{ec para fraccion masica}
\frac{dc_{\vec{u}}}{dt}=-\rho \nabla \cdot \vec{v}_{0}-\nabla \cdot f(\vec{u}-\vec{v}_{0})-\frac{\partial }{\partial \vec{u}}\cdot \vec{J}_{u},
\end{equation}
 obtained by combining the continuity equation (\ref{ec continuidad}) and the
balance of mass (\ref{balance de masa1}),  is now substituted  into
Eq.(\ref{ec. gibbs}). The resulting expression is integrated by parts over
the velocity space assuming that the fluxes vanish at the boundaries. One then
obtains the entropy balance equation in the form

\begin{equation}
\label{balance entropia}
\rho \frac{ds}{dt}=-\nabla \cdot \vec{J}_{s}+\sigma ,
\end{equation}
 where the entropy flux is given by 

\begin{equation}
\label{flujo total de entropia}
\vec{J}_{s}=-k_{B}\int (\vec{u}-\vec{v}_{0})f(lnf-1)d\vec{u}-\frac{m}{2}\int (\vec{u}-\vec{v}_{0})f(\vec{u}-\vec{v}_{0})^{2}d\vec{u},
\end{equation}

\noindent and the entropy production is\begin{equation}
\label{produccion entropia}
\sigma =-\frac{m}{T}\int \vec{J}_{u}\cdot \frac{\partial \mu }{\partial \vec{u}}d\vec{u}-\frac{m}{T}\int \vec{J}\cdot (\vec{u}-\vec{v}_{0})\cdot \nabla \vec{v}_{0}d\vec{u}.
\end{equation}

\noindent This quantity consists of two contributions of the type flux-force
pair: the first one arises from the diffusion process in \( \vec{u} \)-space,
 whereas the second is due to the presence of the  convective
flow  \( \vec{J}\equiv (\vec{u}-\vec{v}_{0})f \). Since both contributions
are vectors, the fluxes couple to the two forces giving rise to cross-effects.
Following the non-equilibrium thermodynamics rules, we can establish linear
phenomenological relationships between fluxes and thermodynamic forces. Assuming
locality in \( \vec{u} \)-space, for which only fluxes and forces with the
same value of \( \vec{u} \) are coupled, the expressions for the currents are
the following

\begin{equation}
\vec{J}_{\vec{u}}=-\frac{m}{T}\vec{\vec{L}}_{\vec{u}\vec{u}}\cdot \frac{\partial \mu }{\partial \vec{u}}-\frac{m}{T}\vec{\vec{L}}_{\vec{u}\vec{r}}\cdot (\vec{u}-\vec{v}_{0})\cdot \nabla \vec{v}_{0},
\end{equation}

\begin{equation}
\label{##}
\vec{J}=-\frac{m}{T}\vec{\vec{L}}_{\vec{r}\vec{r}}\cdot (\vec{u}-\vec{v}_{0})\cdot \nabla \vec{v}_{0}-\frac{m}{T}\vec{\vec{L}}_{\vec{r}\vec{u}}\cdot \frac{\partial \mu }{\partial \vec{u}},
\end{equation}

\noindent where \( \vec{\vec{L}}_{\vec{u}\vec{u}} \), \( \vec{\vec{L}}_{\vec{u}\vec{r}} \),
\( \vec{\vec{L}}_{\vec{r}\vec{u}} \) and \( \vec{\vec{L}}_{\vec{r}\vec{r}} \)
are Onsager coefficients. These coefficients may in general depend on the imposed
velocity gradient. In this case they satisfy generalized Onsager relations
\cite{coeffonsager}.
in which time-reversal symmetry must also be applied to the external driving.
 In the case we are considering, we have \( \vec{\vec{L}}_{\vec{u}\vec{r}}=-\vec{\vec{L}}_{\vec{r}\vec{u}} \).
Defining now the tensors 

\begin{equation}
\label{B}
\vec{\vec{\alpha }}=\frac{m\vec{\vec{L}}_{\vec{u}\vec{u}}}{fT},\; \; \; \; \; \; \vec{\vec{\xi }}=\frac{m\vec{\vec{L}}_{\vec{r}\vec{r}}}{fT},\; \; \; \; \; \; \vec{\vec{\epsilon }}=\frac{m\vec{\vec{L}}_{\vec{u}\vec{r}}}{fT},
\end{equation}

\noindent and using the expression for the chemical potential (\ref{potencial quimico}),
the fluxes can be recast in the form

\begin{equation}
\label{flujo espacio velocidad}
\vec{J}_{\vec{u}}=-(\vec{u}-\vec{v}_{0})\cdot [\vec{\vec{\alpha }}+\nabla \vec{v}_{0}\cdot \vec{\vec{\epsilon }}]f-\frac{k_{B}T}{m}\vec{\vec{\alpha }}\cdot \frac{\partial f}{\partial \vec{u}},
\end{equation}

\begin{equation}
\label{flujo xi}
\vec{J}=-(\vec{u}-\vec{v}_{0})\cdot \left[ \nabla \vec{v}_{0}\cdot \vec{\vec{\xi }}-\vec{\vec{\epsilon }}\right] f+\frac{k_{B}T}{m}\vec{\vec{\epsilon }}\cdot \frac{\partial f}{\partial \vec{u}}.
\end{equation}
 By substituting Eq. (\ref{flujo espacio velocidad}) into the continuity equation
(\ref{ec continuidad}) for the single-particle distribution function, we finally
obtain 

\begin{equation}
\label{fokker-planck general}
\frac{\partial f}{\partial t}+\nabla \cdot \vec{u}f=\frac{\partial }{\partial \vec{u}}\cdot \left\{ (\vec{u}-\vec{v}_{0})\cdot [\vec{\vec{\alpha }}+\nabla \vec{v}_{0}\cdot \vec{\vec{\epsilon }}]f+\frac{k_{B}T}{m}\vec{\vec{\alpha }}\cdot \frac{\partial f}{\partial \vec{u}}\right\} ,
\end{equation}
 which constitutes the Fokker-Planck equation describing the evolution of the
non-equilibrium single-particle distribution function. The fact that the coefficients
appearing in the equation are tensors, reflects the anisotropy of the system
induced by the imposed external flow. Moreover, this equation exhibits the fact
that the flow breaks the Einstein relation by adding a term which depends on
the imposed velocity gradient. This breaking constitutes a proof that the fluctuation-dissipation
theorem cannot be applied when fluctuations take place around the steady state. 

The Fokker-Planck equation we have obtained can be compared with the ones derived
by means of different methods. In Ref. \cite{rodriguez}, authors  found
a similar Fokker-Planck equation for the particular case of a shear flow. In
Section IV, we will discuss that similarity in more detail. Following time-dependent
projector operator techniques, a Fokker-Planck equation similar to ours was
derived in \cite{shea} for the translational modes of a Brownian particles
moving in a flowing bath under a temperature gradient. Since diffusion in a
bath under temperature gradient was studied previously in \cite{temp} in the
framework of MNET, the relevant point to emphasize here is the fact that the
term including the velocity gradient enters the Fokker-Planck equation as an
external force, in a  similar way as in the equation obtained in \cite{shea}.

\section{THE HYDRODYNAMIC EQUATIONS }

In the absence of direct interactions, the single-particle distribution function
provides the complete description of the system at mesoscopic level. Macroscopically,
the description must be carried out in terms of the moments of the distribution
function, which are related to the hydrodynamic fields: density, momentum and
pressure tensor. 

The density defined through Eq. (\ref{definicion densidad}) corresponds to
the zero-order moment. The first-order moment  has been  defined
in (\ref{definicion flujo difusion}). The second moment centered about the
average velocity \( \vec{v} \) of the Brownian `gas' corresponds to the pressure
tensor 

\begin{equation}
\label{definicion tensor de presiones}
\vec{\vec{P}}=m\int (\vec{u}-\vec{v})(\vec{u}-\vec{v})fd\vec{u}.
\end{equation}
 Finally, the third centered moment, which is related to the flux of kinetic
energy and stress, is given by

\begin{equation}
\label{deltaQ}
\vec{\vec{\vec{Q}}}=m\int (\vec{u}-\vec{v})(\vec{u}-\vec{v})(\vec{u}-\vec{v})fd\vec{u}.
\end{equation}

The set of evolution equations for these moments can be obtained by using the
Fokker-Planck equation (\ref{fokker-planck general}) in the definitions of
the conserved quantities, after performing the time derivative and the required
integrations in velocity space  \cite{inertial}. The  evolution
equations for the three first moments are, respectively, the continuity equation

\begin{equation}
\label{balance masa}
\frac{\partial \rho }{\partial t}=-\nabla \cdot \rho \vec{v},
\end{equation}

\noindent the balance of momentum 

\begin{equation}
\label{balance momentum}
\rho \frac{d\vec{v}}{dt}+\nabla \cdot \vec{\vec{P}}=-\rho (\vec{v}-\vec{v}_{0})\cdot \vec{\vec{C}},
\end{equation}

\noindent and the equation for the evolution in time of the pressure tensor 

\begin{eqnarray}
\frac{d}{dt}\vec{\vec{P}}+2(\vec{\vec{P}}\cdot \nabla \vec{v})^{s}+\vec{\vec{P}}\nabla \cdot \vec{v}+2(\vec{\vec{P}}\cdot \vec{\vec{C}})^{s}= &  & \nonumber \\
=\frac{2k_{B}T}{m}\rho \vec{\vec{\alpha }}^{s}-\nabla \cdot \vec{\vec{\vec{Q}}}, & \label{balance tensor presiones} 
\end{eqnarray}
 where \( \vec{\vec{C}}=[\vec{\vec{\alpha }}+\nabla \vec{v}_{0}\cdot \vec{\vec{\epsilon }}] \)
and an upper \( s \) means symmetric part of a tensor. In a similar way, we
could derive the evolution equations for the higher-order moments of the distribution,
which constitute a coupled hierarchy of hydrodynamic equations \cite{inertial,wilemski}. 

Notice that in Eq. (\ref{balance momentum}) the right hand side term can be
identified with the hydrodynamic force exerted by the fluid on the particle,
with \( \vec{\vec{C}} \) playing the role of the friction constant. That friction
constant establishes the characteristic relaxation time scale for the velocity
and it is usually very large. For instance, for a mesoscopic particle of radius
\textbf{\( a\sim 10^{-5}cm \),} moving in a quiescent liquid of viscosity \( \eta \sim 10^{-2}P \),
 the Stokes formula can be applied giving for the friction constant
\textbf{\( \beta =\frac{6\pi \eta a}{m}\sim 10^{8}\: s^{-1} \)}. Consequently,
the discussion of the behavior of the system may be carried out by expanding
the hierarchy of evolution equations for the moments in powers of \( \vec{\vec{C}}^{-1} \).
From the hierarchy of moments, one can easily realize that the ith moment introduces
corrections which are of order \( \vec{\vec{C}}^{-(i-1)} \) for even moments,
and \( \vec{\vec{C}}^{-i} \) for the odd ones. If we drop out the terms arising
from the third and higher moments, we are neglecting corrections to the diffusion
equation of order \( \vec{\vec{C}}^{-3}. \) Notice that in order to retain
\( \vec{\vec{C}}^{-3} \) corrections, one has to include not only the third-order
moment equation, but also the equation for the forth moment  which is
of  the same order. 

The set of equations (\ref{balance masa})-(\ref{balance tensor presiones})
then govern the hydrodynamic behavior of the Brownian `gas' immersed in a fluid
moving with velocity profile \( \vec{v}_{0} \). The elements of the tensor
\( \vec{\vec{C}}^{-1} \)constitute characteristic time scales whose existence
motivates the separation of the dynamics into two well-differentiated regimes:
an inertial regime for \( t\ll (C^{-1})_{ij} \), characterized for the relaxation
of the variables towards the diffusion regime, which is achieved for \( t\gg (C^{-1})_{ij} \)
. Both regimes will be discussed in the following subsections.

\subsection{Inertial regime}

In order to discuss the inertial regime, it is convenient to rewrite Eq. (\ref{balance tensor presiones})
for the evolution of the pressure tensor in the following way

\begin{equation}
\label{relajacion presion}
\frac{1}{2}\frac{d}{dt}\vec{\vec{P}}+\left( \vec{\vec{P}}\cdot \vec{\vec{\tau }}_{1}^{-1}\right) ^{s}=\frac{k_{B}T}{m}\rho \vec{\vec{\alpha }}^{s},
\end{equation}
where we have defined the matrix of relaxation times\[
\vec{\vec{\tau }}_{1}=[\vec{\vec{C}}+\nabla \vec{v}+(\nabla \cdot \vec{v})\vec{\vec{1}}]^{-1},\]
 and we have neglected corrections to the diffusion equation of orders \( \vec{\vec{C}}^{-3} \)
and higher. In an analogous way, from the evolution equation for the momentum,
Eq.(\ref{balance momentum}), we obtain

\begin{equation}
\label{relajacion difusion}
\frac{d\vec{J}_{D}}{dt}+\vec{J}_{D}\cdot \vec{\vec{\tau }}_{2}^{-1}=\rho \vec{v}_{0}\cdot \vec{\vec{C}}-\nabla \cdot \vec{\vec{P}},
\end{equation}
 where \( \vec{J}_{D}\equiv \rho \vec{v} \) and \[
\vec{\vec{\tau }}_{2}=[\vec{\vec{C}}+\vec{\vec{1}}(\nabla \cdot \vec{v})]^{-1},\]
 is the corresponding matrix of relaxation times.

The previous equations describe the inertial regime in the dynamics of the Brownian
particles subjected to stationary flow. This regime holds for times small enough
compared to the characteristic relaxation times, identified with the components
of the matrices \( \vec{\vec{\tau }}_{1} \) and \( \vec{\vec{\tau }}_{2} \). 

Our description of the inertial regime is consistent with the generalized hydrodynamic
description in which the diffusion coefficient depends on the wave vector \cite{boon}.
This property has been shown in \cite{inertial} for the case of a quiescent
liquid.

\subsection{Diffusion regime. }

For times larger than any characteristic relaxation time, \( t\gg (C^{-1})_{ij} \),
the system enters the diffusion regime. In such a regime, the dynamics  becomes
well described by a Smoluchowski equation for the density distribution \( \rho  \)
in the configuration space. 

In order to find the Smoluchowski equation, we will first discuss the diffusion
approximation in the evolution equations for the pressure tensor (\ref{relajacion presion})
and the momentum (\ref{relajacion difusion}). Both equations involve inertial,
friction and velocity gradient time scales.  For \( t\gg (C^{-1})_{ij} \),
time derivatives can be neglected when compared with terms proportional to \( \vec{\vec{C}} \).
Notice also that the term \( \nabla \cdot \vec{v} \) is essentially a time
derivative, as follows from Eq. (\ref{balance de masa1}), and can accordingly
be neglected. Taking this considerations into account, the equations for the
pressure tensor and the momentum then reduce to  

\begin{equation}
\label{ec difusiva P}
(\vec{\vec{P}}\cdot \vec{\vec{C}})^{s}+(\vec{\vec{P}}\cdot \nabla \vec{v})^{s}=\frac{k_{B}T}{m}\rho \vec{\vec{\alpha }}^{s},
\end{equation}
 and 

\begin{equation}
\label{momento difusivo}
\vec{J}_{D}=\rho \vec{v}_{0}-\left( \nabla \cdot \vec{\vec{P}}\right) \cdot \vec{\vec{C}}^{-1},
\end{equation}
respectively. The first equation  represents a linear set of  coupled
 algebraic equations for the components of the pressure tensor \( \vec{\vec{P}} \)
in terms of the components of the tensors \( \vec{\vec{C}} \), \( \nabla \vec{v} \)
and \( \vec{\vec{\alpha }} \). Once the explicit expression for the pressure
tensor has been found, it must be introduced into the momentum equation (\ref{momento difusivo})
yielding the constitutive equation for the diffusion current \( \vec{J}_{D} \).
The Smoluchowski equation is then obtained after substituting \( \vec{J}_{D} \)
into the balance of mass (\ref{balance masa}).

In the following Section we will apply the previous scheme to the particular
case of a shear flow.

\section{DIFFUSION IN A SHEAR FLOW }

Inherent to non-equilibrium thermodynamics is the fact that it cannot provide
explicit expressions for the phenomenological (transport) coefficients which
must be borrowed from other theories. In our case, the still unspecified quantities
are \( \vec{\vec{\alpha }} \) and \( \vec{\vec{\epsilon }} \). When a specific
velocity profile is given, expressions for those tensors can be obtained from
kinetic theory or hydrodynamics. 

The formalism developed in the previous sections is valid for an arbitrary stationary
velocity field. In this section we will focus our discussion on the particular
case of Brownian motion in a shear flow. Our first task will be the identification
of the phenomenological coefficients. As pointed out before, the second term
on the right hand side of equation (\ref{balance momentum}) can be identified
with the force per unit of mass exerted on a suspended particle by the host
fluid, where \( \vec{\vec{C}} \) plays the role of the friction tensor. In
general, such a density force can be calculated from hydrodynamics, (see for
instance Refs. \cite{saffman,bedaux}). For the shear flow case, the friction
coefficient contains in general linear and quadratic contributions in \( \zeta a \),
where \( \zeta  \) is the inverse penetration length of the perturbations \( \zeta =\left( \frac{\gamma }{\nu }\right) ^{\frac{1}{2}}, \)
with \( \gamma  \) being the shear rate and \( \nu  \) the kinematic viscosity.
For a Brownian particle under moderate shear flow the term \( \zeta a \) is
very small, consequently the friction tensor can be approximated by \( \vec{\vec{C}}\simeq \beta \vec{\vec{1}} \).
This identification leads to the following expression for the tensor \( \vec{\vec{\alpha }} \),

\begin{equation}
\label{B shear flow}
\vec{\vec{\alpha }}=\beta \vec{\vec{1}}-\nabla \vec{v}_{0}\cdot \vec{\vec{\epsilon }}.
\end{equation}

\noindent By introducing the previous expressions into the Fokker-Planck equation
(\ref{fokker-planck general}), we finally obtain 

\begin{equation}
\label{fp para shear}
\frac{\partial f}{\partial t}+\vec{u}\cdot \nabla f=\frac{\partial }{\partial \vec{u}}\cdot \left\{ \beta (\vec{u}-\vec{v}_{0})f+\frac{k_{B}T}{m}\vec{\vec{\alpha }}\cdot \frac{\partial f}{\partial \vec{u}}\right\} .
\end{equation}
 As concluded also from Eq.(\ref{fokker-planck general}), the particular form
of Eq.(\ref{fp para shear}) implies that the fluctuation-dissipation theorem
is no longer valid when the fluid (heat bath) is sheared. Notice that the term
which invalidates that theorem is proportional to the velocity gradient, or
to the inverse penetration length squared. The theorem remains applicable in
the case in which the friction coefficient contains a correction proportional
to \( \zeta  \) \cite{zetarubi}.

The Fokker-Planck equation (\ref{fp para shear}) is similar to the corresponding
one obtained in Ref. \cite{rodriguez} from kinetic theory. By expanding the
collision operator in the mass ratio between the fluid and Brownian particles,
authors derived a Fokker-Planck equation in which the diffusion tensor contains
a correction to the Einstein formula proportional to the stress tensor of the
fluid. Since this quantity is proportional to the velocity gradient we conclude
that the form of the diffusion tensor they find is similar to our expression
(\ref{B shear flow}), which allow the identification of the tensor \textbf{\( \vec{\vec{\epsilon }} \).}

With the explicit form of the Fokker-Planck equation (\ref{fp para shear})
in mind, our purpose is now to discuss the macroscopic evolution of the system.
Following the procedure indicated in subsection  III.B, we find that
for a shear flow, in the diffusion regime, the expression for the Brownian pressure
tensor is

\begin{equation}
\label{tensor de presiones sin inerciales}
\vec{\vec{P}}=\frac{k_{B}T}{m}\rho \left[ \vec{\vec{1}}-\left( \beta ^{-1}(\vec{\vec{1}}+\vec{\vec{\epsilon }})\cdot \nabla \vec{v}_{0}\right) ^{s}\right] .
\end{equation}
 From this equation,  we can conclude that Brownian  motion
of the particles contributes to the total pressure tensor of the suspension
in two forms. The first contribution is the well-known scalar kinetic pressure
given by

\begin{equation}
\label{ec-estado}
p=\frac{k_{B}T}{m}\rho ,
\end{equation}
 which is the equation of state for the ideal Brownian `gas'. The second contribution
comes from the irreversible part \( \vec{\vec{\Pi }} \) of the Brownian pressure
tensor, which can be written in the form

\begin{equation}
\label{pi griega brownies}
\vec{\vec{\Pi }}=-D_{0}\rho \left[ (\vec{\vec{1}}+\vec{\vec{\epsilon }})\cdot \nabla \vec{v}_{0}\right] ^{s},
\end{equation}
where \( D_{0}=\frac{k_{B}T}{m\beta } \) is the diffusion coefficient of a
particle when the liquid is at rest. This last equation defines the Brownian
viscosity tensor 

\begin{equation}
\label{viscosidad}
\vec{\vec{\eta }}_{B}=D_{0}\rho (\vec{\vec{1}}+\vec{\vec{\epsilon }}),
\end{equation}
 which contains the `Brownian' viscosity \( D_{0}\rho  \) \cite{freed}, and
the contribution due to  the coupling with the non-equilibrium bath
which is proportional to \( \vec{\vec{\epsilon }} \). Eqs. (\ref{pi griega brownies})
and (\ref{viscosidad}) accounts for the contributions to the irreversible part
of the pressure tensor and shear viscosity coefficient of the suspension. No
contribution to the bulk viscosity has been found since the shear flow is incompressible.

Following the steps described previously, we can obtain the diffusion current
\( \vec{J}_{D} \) in the limit \( t\gg (C^{-1})_{ij} \), in which Eq.(\ref{relajacion difusion})
has the form

\begin{equation}
\label{Jd en regimen difusivo}
\vec{J}_{D}=\rho \vec{v}_{0}-\beta ^{-1}\nabla \cdot \vec{\vec{P}}.
\end{equation}
Introducing  the expression for the pressure tensor (\ref{tensor de presiones sin inerciales})
into the last equation we obtain,  

\begin{equation}
\label{Jd alfa}
\vec{J}_{D}=\rho \vec{v}_{0}-\vec{\vec{D}}\cdot \nabla \rho -\rho \nabla \cdot \vec{\vec{D}}.
\end{equation}
 Here we have defined the diffusion tensor as 

\begin{equation}
\label{tensor de difusion}
\vec{\vec{D}}=D_{0}\left[ \vec{\vec{1}}-\beta ^{-1}\left( (\vec{\vec{1}}+\vec{\vec{\epsilon }})\cdot \nabla \vec{v}_{0}\right) ^{s}\right] .
\end{equation}
 An important consequence of this result is that the presence of a shear flow
modifies the diffusion current with respect to the case of a quiescent liquid.
Substituting Eq.(\ref{Jd alfa}) into the balance of mass, Eq.(\ref{balance masa}),
we finally obtain the diffusion equation for the Brownian particle 

\begin{equation}
\label{ec.  Smoluchowski}
\frac{\partial \rho }{\partial t}=-\nabla \cdot \rho \vec{v}_{0}+\nabla \cdot (\vec{\vec{D}}\cdot \nabla \rho )+\nabla \cdot (\rho \nabla \cdot \vec{\vec{D}}).
\end{equation}

This equation is a generalization of the usual Smoluchowski equation, in the
sense that it contains the anisotropic diffusion coefficient (\ref{tensor de difusion})
which depends on the imposed velocity gradient. For the particular case of the
shear flow, the velocity gradient is a constant, therefore \( \vec{\vec{D}} \)
does not exhibit  spatial dependence and we recover the usual Smoluchowski
equation \begin{equation}
\label{ec.  Smoluchowski final}
\frac{\partial \rho }{\partial t}=-\nabla \cdot \rho \vec{v}_{0}+\nabla \cdot (\vec{\vec{D}}\cdot \nabla \rho ).
\end{equation}
 This equation coincides formally with the one found in Ref.  \cite{feldi 2}.
The difference between both equations lies in the form of the diffusion tensor.
In the cited work, author starts from a Fokker-Planck equation which satisfies
the fluctuation-dissipation theorem. In the diffusion regime author finds, consistently
with the approximation, that the diffusion tensor and the mobility of the Brownian
particles in the fluid are related each other through Einstein's formula.

\section{DISCUSSION}

In this paper we have analyzed the dynamics of a suspension of Brownian particles
in a non-equilibrium situation resulting from the action of an externally imposed
velocity gradient. We have applied the method of mesoscopic non-equilibrium
thermodynamics to study the dissipation in phase space related to the underlying
diffusion process of the probability density of the particles.

In MNET, local equilibrium is assumed at mesoscopic level. A Gibbs equation
is then proposed in which the entropy depends on the probability density: the
pertinent density in phase space, according to the concept of Gibbs entropy.
By applying the rules of non-equilibrium thermodynamics, one obtains the entropy
production of the system, which enables one to derive the expression for the
diffusion current in phase space and, consequently, to obtain the Fokker-Planck
equation for the single-particle distribution function. This expression exhibits
violation of the fluctuation-dissipation theorem whose origin is precisely the
presence of the external gradient. This feature, commonly found in the wide
class of driven-diffusion systems \cite{lesbowitz,zetarubi}, has also been
reported in slow relaxation process of glassy systems \cite{kurchan}. 

The hydrodynamic level of description is accomplished from the evolution equations
for the first moments of the distribution function, which can be obtained through
the Fokker-Planck equation. The time evolution of the moments include relaxation
equations for the diffusion current and the pressure tensor, whose form permits
to elucidate the existence of inertial  (short-time) and diffusion (long-time)
regimes. In the diffusion regime, the mesoscopic description is carried out
by means of a Smoluchowski equation. In this regime\textbf{,} the equations
for the moments coincide with the differential equations of non-equilibrium
thermodynamics. The equations for the moments obtained from the Fokker-Planck
equation in the framework of MNET extend the domain of applicability of non-equilibrium
thermodynamics to shorter time scales. These equations can be reformulated in
terms of transport coefficients which depends on the wave vector, in accordance
with generalized hydrodynamics \cite{boon}.

Our results can be compared with others obtained previously using different
theories. The Fokker-Planck equation we have derived is similar to the one proposed
in \cite{rodriguez} for the case of a shear flow. As in our case, these authors
show that when the system is sheared, the diffusion coefficient is no longer
given by Stokes-Einstein law. It contains a correction proportional to the pressure
tensor which is basically of the same nature as the one we obtain, i.e., proportional
to the velocity gradient. This fact clearly shows violation of the fluctuation-dissipation
theorem due to the presence of the external input necessary to maintain the
stationary state. Similar conclusions are  obtained in \cite{shea},
where the general case in which a stationary flow and a temperature gradients
act simultaneously is studied by means of  time-dependent projector
operator techniques. 

Having discussed the comparison of our results with others coming from statistical
mechanical theories, it remains to analyze them in the framework of thermodynamical
theories dealing with systems outside equilibrium, in particular with extended
 irreversible  thermodynamics. This theory provides hydrodynamic
equations which also  contain relaxation terms for the pressure tensor
and the diffusion current\textbf{.} Along our analysis of this and other cases
of systems outside equilibrium (see Refs. \cite{n-particles,inertial,fluctuations,preprint}),
we have shown that by simply using non-equilibrium thermodynamics or its extension
to the mesoscopic domain (MNET), we are able to completely characterize the
evolution of systems outside equilibrium. This fact question the need of using
generalized entropies \cite{extended} depending on non-thermodynamic variables:
the fluxes, which constitutes the cornerstone of extended irreversible thermodynamics.
 The application of the well-established non-equilibrium thermodynamics
postulates as indicated in \cite{degroot}, suffices to provide a general scheme
under which non-equilibrium processes of macroscopic and mesoscopic nature can
be treated.

Far from being specific of the case of a suspension of non-interacting Brownian
particles, we have treated in this paper, the method we have presented could
systematically be applied to analyze the dynamics of soft condensed matter systems
under shear flow. For example, the case in which direct and hydrodynamic interactions
among particles become  important  can be worked out along the
lines indicated in \cite{n-particles}, for the case of a quiescent liquid.
If the suspended objects are deformable or need of additional parameters to
characterize their state, as may occur in polymers \cite{Doi} or liquid crystals
\cite{preprint,gennes}, the set of variables of the distribution function has
to be enlarged. One would obtain the corresponding Fokker-Planck equation and
from it the evolution equations for the moments which define the hydrodynamic
or generalized hydrodynamic regimes. In all these cases, kinetic and hydrodynamic
equations can be derived following the method of mesoscopic non-equilibrium
thermodynamics.

\section{ACKNOWLEDGMENTS}

I. S. H. wants to acknowledge J. Dufty and A. P\'{e}rez Madrid for their comments,
and to UNAM-DGAPA for economic support. D. Reguera wishes to thank Generalitat
de Catalunya for financial support. This work has been partially supported by
DGICYT of the Spanish Government under grant PB98-1258.


\begin{thebibliography}{10}
\bibitem{lenan}J. A. McLennan, \textit{Introduction to Non-equilibrium Statistical Mechanics}
(Prentice Hall, 1989). 
\bibitem{libro-keizer}J. Keizer, \textit{Statistical Thermodynamics of Nonequilibrium Processes} (Springer-Verlag,
New York, 1987).
\bibitem{schmitz}R. Schmitz, Phys. Rep. \textbf{171}, 1 (1988).
\bibitem{lesbowitz}G.L. Eyink, J. L. Lebowitz, H. Spohn, J. Stat. Phys. \textbf{83}, 385 (1996). 
\bibitem{dorfman}J. R. Dorfman, T. R. Kirkpatrick, J. V. Sengers, Ann. Rev. Phys. Chem. \textbf{45},
213 (1994).
\bibitem{long-range-correl}I. Pagonabarraga, J. M. Rub\'{\i}, Phys. Rev. E \textbf{49}, 267 (1994). 
\bibitem{onuki}A. Onuki, J. Phys.: Condens. Matter \textbf{9} 6119 (1997).
\bibitem{fragile}M. E. Cates, M. R. Evans (Eds.), \textit{Soft and Fragile Matter} (Institute
of Physics Publishing, Bristol, 2000).
\bibitem{Doi}M. Doi, S. F. Edwards, \textit{The Theory of Polymer Dynamics} (Oxford University
Press, New York, 1988).
\bibitem{vandeven}T. G. M. van de Ven, \textit{Colloidal Hydrodynamics} (Academic Press 1989).
\bibitem{vilar}J. M. Rub\'{\i}, J. M. G. Vilar, J. Phys. Condens. Matter \textbf{12},  A75
(2000) .
\bibitem{jaeger}H. M. Jaeger, S. R. Nagel, R. P. Behringer, Rev. Mod. Phys. \textbf{68}, 1259
(1996).
\bibitem{chen}Z. R. Chen et al., Science \textbf{277}, 1248 (1997).
\bibitem{liu}C. Liu, D. J. Pine, Phys. Rev. Lett. \textbf{77}, 2121 (1996).
\bibitem{yamamoto}J. Yamamoto, H. Tanaka, Phys. Rev. Lett. \textbf{77}, 4390 (1996).
\bibitem{physicalkietics}E. M. Lifshitz, L. P. Pitaevskii, \textit{Physical Kinetics,} (Pergamon Press,
1981).
\bibitem{Zubarev}D. N. Zubarev, \textit{Non-equilibrium Statistical Thermodynamics} (Consultants
Bureau, New York-London, 1974).
\bibitem{kampen}N. G. van Kampen, \textit{Stochastic Processes in Physics and Chemistry} (North
Holland, Amsterdam, 1992).
\bibitem{kampi-oppi}N. G. van Kampen, I. Oppenheim, Physica A \textbf{138}\textit{,} 231 (1986).
\bibitem{deutch}J. M. Deutch, I. Oppenheim, Faraday Disc. Chem. Soc. \textbf{83}, 1 (1987).
\bibitem{degroot}S. R. de Groot, P. Mazur, \textit{Non-equilibrium Thermodynamics} (Dover, New
York, 1984). 
\bibitem{temp}A. P\'{e}rez-Madrid, J. M. Rub\'{\i}, P. Mazur, Physica A \textbf{212}, 231
(1994). 
\bibitem{ignacio}I. Pagonabarraga, A. P\'{e}rez-Madrid, J. M. Rub\'{\i}, Physica A \textbf{237},
205 (1997).
\bibitem{n-particles}J. M. Rub\'{\i}, P. Mazur, Physica A \textbf{250}, 253 (1998).
\bibitem{inertial}J. M. Rub\'{\i}, A. P\'{e}rez-Madrid, Physica A \textbf{264}, 492 (1999). 
\bibitem{prefluctuations}P. Mazur, Physica A \textbf{261}, 451 (1998).
\bibitem{fluctuations}P. Mazur, Physica A \textbf{274}, 491 (1999).
\bibitem{hydrodynamic-fluctuations}J. M. Rub\'{\i}, P. Mazur, Physica A \textbf{276}, 477 (2000).
\bibitem{rodriguez}R. Rodriguez, E. Salinas-Rodriguez, J. Dufty, J. Stat. Phys. \textbf{32}, 279
(1983). 
\bibitem{hover}W. T. Ashurst, W. G. Hoover, Phys. Rev. A \textbf{11}, 658 (1975).
\bibitem{keizer}E. Peakock-Lopez, J. Keizer, Phys. Lett. \textbf{108}, 85 (1985).
\bibitem{coeffonsager}J. W. Dufty, J. M. Rub\'{\i}, Phys. Rev. A \textbf{36}, 222 (1987).
\bibitem{shea}J. E. Shea and I. Oppenheim, Physica A \textbf{250}, 265 (1998).
\bibitem{wilemski}G. Wilemski, J. Stat. Phys. \textbf{14}, 153 (1976).
\bibitem{boon}J. P. Boon, S. Yip, \textit{Molecular Hydrodynamics} (Dover, New York, 1980).
\bibitem{saffman}P. G. Saffman, J. Fluid Mech. \textbf{22} 385 (1965).
\bibitem{bedaux}A. P\'{e}rez-Madrid, J. M. Rub\'{\i}, D. Bedeaux, Physica A \textbf{163},
778 (1990).
\bibitem{zetarubi}J. M. Rub\'{\i}, D. Bedeaux, J. Stat. Phys. \textbf{53}, 125 (1988).
\bibitem{freed}K. F. Freed and M. Muthukumar, J. Chem. Phys. \textbf{69}, 2657 (1978). 
\bibitem{feldi 2}B. U. Felderhof, Physica A \textbf{147}, 203 (1987).
\bibitem{kurchan}J.P. Bouchaud, L.F. Cugliandolo, J. Kurchan, and M. Mézard in \emph{Spin Glasses
and Random Fields} (A.P. Young Ed., World Scientific, Singapore, 1997).
\bibitem{preprint}D. Bedeaux, J. M. Rub\'{\i}, \textit{}(preprint)\textit{.}
\bibitem{extended}D. Jou, J. Casas-Vazquez, G. Lebon, Rep. Prog. Physics \textbf{62}, 1035 (1999). 
\bibitem{gennes}P. G. de Gennes, \textit{The Physics of Liquid Crystals} (Clarendon Press, Oxford,
1974).\end{thebibliography}
\end{document}